\documentclass[12pt]{article}
\usepackage{amssymb, amsmath}

\newcommand{\ra}{\rightarrow}
\newcommand{\lra}{\leftrightarrow}

\def\Equiv{\Longleftrightarrow}

\def\PCR{{\bf PC(R)}}
\def\PqR{\Pi^q_1(R)}
\def\NP{{\bf NP}}
\def\PKR{{\bf PK(R)}}
\def\PK{{\bf PK}}
\def\QPCR{{\bf QPC(R)}}
\def\QPC{{\bf QPC}}
\def\GR{{\bf G(R)}}
\def\GGstarR{{\bf G_2^*(R)}}
\def\GstR{{\bf G_1^*(R)}}
\def\G{{\bf G}}
\def\LK{{\bf LK}}
\def\NEXP{{\bf NEXP}}
\def\AX{{\bf AX}}
\def\WPHPR{{\bf WPHP}(R,n)}
\def\PHPR{{\bf PHP}(R,n)}
\def\PHPa{{\bf PHP}^{a^2}_a}
\def\vp{\vec{p}}
\def\vq{\vec{q}}
\def\vg{\vec{g}}
\def\vr{\vec{r}}
\def\vs{\vec{s}}

\newdimen\proofrulebreadth \proofrulebreadth=.05em
\newdimen\proofdotseparation \proofdotseparation=1.25ex
\newdimen\proofrulebaseline \proofrulebaseline=2ex
\newcount\proofdotnumber \proofdotnumber=3
\let\then\relax
\def\hfi{\hskip0pt plus.0001fil}
\mathchardef\squigto="3A3B
\newif\ifinsideprooftree\insideprooftreefalse
\newif\ifonleftofproofrule\onleftofproofrulefalse
\newif\ifproofdots\proofdotsfalse
\newif\ifdoubleproof\doubleprooffalse
\let\wereinproofbit\relax
\newdimen\shortenproofleft
\newdimen\shortenproofright
\newdimen\proofbelowshift
\newbox\proofabove
\newbox\proofbelow
\newbox\proofrulename
\def\shiftproofbelow{\let\next\relax\afterassignment\setshiftproofbelow\dimen0 }
\def\shiftproofbelowneg{\def\next{\multiply\dimen0 by-1 }%
\afterassignment\setshiftproofbelow\dimen0 }
\def\setshiftproofbelow{\next\proofbelowshift=\dimen0 }
\def\setproofrulebreadth{\proofrulebreadth}

\def\prooftree{
\ifnum  \lastpenalty=1
\then   \unpenalty
\else   \onleftofproofrulefalse
\fi
\ifonleftofproofrule
\else   \ifinsideprooftree
        \then   \hskip.5em plus1fil
        \fi
\fi
\bgroup
\setbox\proofbelow=\hbox{}\setbox\proofrulename=\hbox{}%
\let\justifies\proofover\let\leadsto\proofoverdots\let\Justifies\proofoverdbl
\let\using\proofusing\let\[\prooftree
\ifinsideprooftree\let\]\endprooftree\fi
\proofdotsfalse\doubleprooffalse
\let\thickness\setproofrulebreadth
\let\shiftright\shiftproofbelow \let\shift\shiftproofbelow
\let\shiftleft\shiftproofbelowneg
\let\ifwasinsideprooftree\ifinsideprooftree
\insideprooftreetrue
\setbox\proofabove=\hbox\bgroup$\displaystyle 
\let\wereinproofbit\prooftree
\shortenproofleft=0pt \shortenproofright=0pt \proofbelowshift=0pt
\onleftofproofruletrue\penalty1
}

\def\eproofbit{
\ifx    \wereinproofbit\prooftree
\then   \ifcase \lastpenalty
        \then   \shortenproofright=0pt  
        \or     \unpenalty\hfil         
        \or     \unpenalty\unskip       
        \else   \shortenproofright=0pt  
        \fi
\fi
\global\dimen0=\shortenproofleft
\global\dimen1=\shortenproofright
\global\dimen2=\proofrulebreadth
\global\dimen3=\proofbelowshift
\global\dimen4=\proofdotseparation
\global\count255=\proofdotnumber
$\egroup  
\shortenproofleft=\dimen0
\shortenproofright=\dimen1
\proofrulebreadth=\dimen2
\proofbelowshift=\dimen3
\proofdotseparation=\dimen4
\proofdotnumber=\count255
}

\def\proofover{
\eproofbit 
\setbox\proofbelow=\hbox\bgroup 
\let\wereinproofbit\proofover
$\displaystyle
}%
\def\proofoverdbl{
\eproofbit 
\doubleprooftrue
\setbox\proofbelow=\hbox\bgroup 
\let\wereinproofbit\proofoverdbl
$\displaystyle
}%
\def\proofoverdots{
\eproofbit 
\proofdotstrue
\setbox\proofbelow=\hbox\bgroup 
\let\wereinproofbit\proofoverdots
$\displaystyle
}%
\def\proofusing{
\eproofbit 
\setbox\proofrulename=\hbox\bgroup 
\let\wereinproofbit\proofusing
\kern0.3em$
}

\def\endprooftree{
\eproofbit 
  \dimen5 =0pt
\dimen0=\wd\proofabove \advance\dimen0-\shortenproofleft
\advance\dimen0-\shortenproofright
\dimen1=.5\dimen0 \advance\dimen1-.5\wd\proofbelow
\dimen4=\dimen1
\advance\dimen1\proofbelowshift \advance\dimen4-\proofbelowshift
\ifdim  \dimen1<0pt
\then   \advance\shortenproofleft\dimen1
        \advance\dimen0-\dimen1
        \dimen1=0pt
        \ifdim  \shortenproofleft<0pt
        \then   \setbox\proofabove=\hbox{%
                        \kern-\shortenproofleft\unhbox\proofabove}%
                \shortenproofleft=0pt
        \fi
\fi
\ifdim  \dimen4<0pt
\then   \advance\shortenproofright\dimen4
        \advance\dimen0-\dimen4
        \dimen4=0pt
\fi
\ifdim  \shortenproofright<\wd\proofrulename
\then   \shortenproofright=\wd\proofrulename
\fi
\dimen2=\shortenproofleft \advance\dimen2 by\dimen1
\dimen3=\shortenproofright\advance\dimen3 by\dimen4
\ifproofdots
\then
        \dimen6=\shortenproofleft \advance\dimen6 .5\dimen0
        \setbox1=\vbox to\proofdotseparation{\vss\hbox{$\cdot$}\vss}%
        \setbox0=\hbox{%
                \advance\dimen6-.5\wd1
                \kern\dimen6
                $\vcenter to\proofdotnumber\proofdotseparation
                        {\leaders\box1\vfill}$%
                \unhbox\proofrulename}%
\else   \dimen6=\fontdimen22\the\textfont2 
        \dimen7=\dimen6
        \advance\dimen6by.5\proofrulebreadth
        \advance\dimen7by-.5\proofrulebreadth
        \setbox0=\hbox{%
                \kern\shortenproofleft
                \ifdoubleproof
                \then   \hbox to\dimen0{%
                        $\mathsurround0pt\mathord=\mkern-6mu%
                        \cleaders\hbox{$\mkern-2mu=\mkern-2mu$}\hfill
                        \mkern-6mu\mathord=$}%
                \else   \vrule height\dimen6 depth-\dimen7 width\dimen0
                \fi
                \unhbox\proofrulename}%
        \ht0=\dimen6 \dp0=-\dimen7
\fi
\let\doll\relax
\ifwasinsideprooftree
\then   \let\VBOX\vbox
\else   \ifmmode\else$\let\doll=$\fi
        \let\VBOX\vcenter
\fi
\VBOX   {\baselineskip\proofrulebaseline \lineskip.2ex
        \expandafter\lineskiplimit\ifproofdots0ex\else-0.6ex\fi
        \hbox   spread\dimen5   {\hfi\unhbox\proofabove\hfi}%
        \hbox{\box0}%
        \hbox   {\kern\dimen2 \box\proofbelow}}\doll%
\global\dimen2=\dimen2
\global\dimen3=\dimen3
\egroup 
\ifonleftofproofrule
\then   \shortenproofleft=\dimen2
\fi
\shortenproofright=\dimen3
\onleftofproofrulefalse
\ifinsideprooftree
\then   \hskip.5em plus 1fil \penalty2
\fi
}

\begin{document}

\begin{center}
{\large\bf Relativized Propositional Calculus}\\
Stephen Cook\\
Working paper, March, 2012\footnote{This is a slight revision of
a working paper from June 4, 2003.  Much of this material was
presented at the complexity theory workshop at Overwolfach,
30 April, 2003.}
\end{center}

{\bf Motivation:}\\
Complexity lower bounds and independence results are
easier in a relativized setting.  It seems reasonable to define a
relativized setting for the propositional calculus in order to prove
lower bounds.

{\bf Syntax:}\\
The language of {\bf PC(R)}  (propositional calculus relativized to R)
consists of formulas built from atoms $p,q,r,...$ using
the usual connectives 0,1,$\wedge,\vee,\neg$, together with the
relation symbol $R$.  The usual formation rules for formulas apply,
but in addition we agree that
$$\mbox{if $A_1,...,A_n$ are formulas, $n\geq 0$, then $R(A_1,...,A_n)$
is a formula}  $$

{\bf Semantics:}  \\
A structure $\tau$ consists of an assignment of a truth value $p^\tau$
in $\{0,1\}$ (where 1 = TRUE and 0 = FALSE)
to each atom $p$, together with a set $R^\tau\subseteq \{0,1\}^*$
of binary strings.  Then each formula $A$ of \PCR gets a truth
value $A^\tau \in \{0,1\}$ in the obvious way.  In particular, 
$$  R(A_1,...,A_n)^\tau = 1 \Equiv A_1^\tau...A_n^\tau \in R^\tau  $$

This syntax and semantics is essentially the same as that defined
by Ben-David and Gringauze [3].  (See also [1].)

We say that $A$ is {\em valid} iff $A^\tau = 1$ for all structures $\tau$,
and $A$ is {\em satisfiable} iff $A^\tau=1$ for some structure $\tau$.

For example,
$$  (R(p)\wedge R(\neg p))\supset (R(q)\vee R(\neg q))  $$
is valid.  In general, $A$ is valid iff $\neg A$ is unsatisfiable.

{\bf Theorem 1:}  The satisfiability problem for \PCR formulas is in
\NP\ (and hence \NP-complete).

{\bf Proof:}  A certificate for satisfiability need only specify
$\tau$ for each atom in $A$, and for each occurrence of the form
$R(B_1,...,B_n)$ in $A$, some string $v_1...v_n\in\{0,1\}^n$ is
specified to either be in $R^\tau$ or not in $R^\tau$.  $\Box$

{\bf System PK(R):}\\
This is Gentzen's sequent system {\bf PK} for the propositional
calculus (see for example [4] or [5]), except formulas are allowed to be
\PCR formulas, and in addition to the axiom scheme $A\ra A$,
and the axioms $\ra 1$ and $0\ra$,
we add the axiom scheme

${\bf AX:} \qquad \neg A\vee B, A\vee \neg B, R(\vec{C},A,\vec{D})\ra R(\vec{C},B,\vec{D})$

which asserts that if $A$ and $B$ are equivalent, then one can be
substituted for the other as an argument of $R$.

Using \AX, each of the following four schemes E1,E2,E3,E4
has a \PKR\ proof with a constant number of sequents:

E1) $A,R(\vec{C},A,\vec{D}) \ra R(\vec{C},1,\vec{D})$\\
E2) $A,R(\vec{C},1,\vec{D}) \ra R(\vec{C},A,\vec{D})$\\
E3) $R(\vec{C},A,\vec{D}) \ra A, R(\vec{C},0,\vec{D})$\\
E4) $R(\vec{C},0,\vec{D}) \ra A, R(\vec{C},A,\vec{D})$

{\bf Theorem 2:}  \PKR\ is sound and complete.  Further every valid
sequent $S$ has a \PKR\ proof $\pi$ with $O(2^{|S|})$ sequents, where 
each sequent in $\pi$ has length $O(|S|)$, where $|S|$ is the total number
of symbols in $S$.

{\bf Remark:} In counting the number of sequents in a proof, we
do not count weakenings and exchanges.

{\bf Proof:}  Soundness asserts that every sequent derivable in \PKR\
is valid.  This is true because the axioms are valid and the rules
preserve validity.

Completeness asserts that every valid sequent has a \PKR\ proof.
To get an upper bound on the number of lines in the proof, we
make the following definition:

{\bf Definition 1:}  The cost $c(A)$ of a formula $A$ is the number
of occurrences of $\wedge,\vee,\neg$ in $A$ plus, for each subformula
$R(B_1,...,B_n)$ in $A$, the number of formulas in the sequence
$B_1,...,B_n$ other than 0 or 1.   The cost $c(S)$ of a sequent $S$ is the
sum of the costs of the formulas in the sequent.

For example, the cost of $R(p\wedge q, p\wedge q, p ,0,1,1)$ is 5:
2 for the two occurrences of $\wedge$, and 3 for the three nontrivial
arguments of $R$.

Note that $c(A)\leq |A|$, where $|A|$ is the number of symbols in $A$,
counting commas.

{\bf Lemma 1:} For some constant $d$, each valid sequent $S$ has a
\PKR\ proof with at most $d2^{c(S)}$ lines,
where each line has length $O(|S|)$.

{\bf Proof:}  Induction on $c(S)$.

{\bf Basis:}  Suppose that $c(\Gamma\ra\Delta)=0$ and $\Gamma\ra\Delta$ is valid.
Then any occurrence of $R$ must be as a formula of the form $R(v_1,...,v_n)$
in one of the sequences $\Gamma$ or $\Delta$,
where each $v_i$  is either 0 or 1.  It is easy to check that
either $1$ is a formula in $\Delta$, or 0 is a
formula in $\Gamma$, or some formula occurs in both
$\Gamma$ and $\Delta$.  In each case, $\Gamma\ra\Delta$ can be
derived from an axiom (other than \AX)) by weakenings
and exchanges.

{\bf Induction Step:}  $c(\Gamma\ra\Delta)>0$.  Then some formula
in either $\Gamma$ or $\Delta$ must have a principal connective that
is either $\wedge,\vee,\neg$ or $R$.  For the cases $\wedge,\vee, \neg$
we derive $\Gamma\ra\Delta$ by the appropriate \PK\ introduction
rule (left or right), thus reducing the problem to deriving one
or two valid sequents, each of reduced cost, so the Induction Hypothesis
applies.

Now suppose that $\Gamma\ra\Delta$ has the form
\begin{equation}\label{form}
   \Gamma'\ra\Delta', R(\vec{C},A,\vec{D})
\end{equation}
where $A$ is not 0 or 1.  Then we use the derivation below,
based on E2 and E4 above,
where we have omitted weakenings  and exchanges.  All indicated inferences
use the {\bf cut} rule.
$$
\begin{prooftree}
   \[  A,\Gamma'\ra\Delta', R(\vec{C},1,\vec{D}) \qquad  E2
    \justifies A,\Gamma'\ra\Delta', R(\vec{C},A,\vec{D})  \]
  \ \ 
  \[  \Gamma'\ra\Delta',A, R(\vec{C},0,\vec{D})  \qquad E4
     \justifies \Gamma'\ra\Delta',A, R(\vec{C},A,\vec{D}) \]
     \justifies \Gamma'\ra\Delta', R(\vec{C},A,\vec{D})
\end{prooftree}
$$
This reduces the proof of (\ref{form}) to the proof of two valid sequents,
each of which has cost one less than the cost of (\ref{form}).
The induction hyposthesis applied to these two sequents gives
us the desired result.

The remaining case to consider is that $\Gamma\ra\Delta$ has the
form
$$ \Gamma',R(\vec{C},A,\vec{D})\ra\Delta'  $$
where again $A$ is not 0 or 1.  This time we use the derivation
below, using E1 and E3:
$$
\begin{prooftree}
   \[  A, R(\vec{C},1,\vec{D}),\Gamma'\ra\Delta' \qquad E1
    \justifies A, R(\vec{C},A,\vec{D}),\Gamma'\ra\Delta' \]
  \ \
  \[ R(\vec{C},0,\vec{D}), \Gamma'\ra\Delta',A  \qquad E3
     \justifies R(\vec{C},A,\vec{D}),\Gamma'\ra \Delta',A \]
     \justifies \Gamma',R(\vec{C},A,\vec{D})\ra\Delta'
\end{prooftree}
$$
Now we apply the induction hypothesis, as in the  previous case.

\begin{center}
{\large\bf Quantified Relativized Propositional Calculus}
\end{center}

Formulas in \QPCR\ are like those in \PCR, except we now allow
quantifiers $\forall x$ and $\exists x$, for an atom $x$.
The semantics are obtained in the obvious way by letting $x$ range
over $\{0,1\}$.

{\bf Notation:} $\Pi^q_1(R)$ is the class of formulas of \QPCR\
of the form
$$      \forall \vec{x} A(\vec{x}, \vec{p}, R)  $$ 
where $A$ is quantifier-free.

{\bf Theorem 3:}  The satisfiability problem for \QPCR\ is complete
for \NEXP.  The same is true for the satisfiability problem
restricted to $\Pi^q_1(R)$ formulas.

{\bf Proof:}  It is easy to see that the satisfiability problem
is in \NEXP:  Given a formula $A$ of \QPCR, let $n$ be the largest
number of arguments of any occurrence if $R$ in $A$.
Guess at a structure $\tau$ for $A$ by writing down truth values
to the free variables of $A$, and specifying $R^\tau$ for $R$ up to
strings of length $n$ by writing down a subset of $\{0,1\}^{\leq n}$.  
Now verify that $\tau$ satisfies $A$.

Hardness can be established either by a direct reduction of
Turing machine computations to \QPCR\ satisfiability
(proof due to Charles Rackoff), or
by using the proof that succint circuit satisfiability is NEXP
complete (see page 494 of Christos Papadimitriou's textbook
on Computational Complexity) (proof due to Tsuyoshi Morioka).

{\bf Notation:}  $|A|$ denotes the length of a formula $A$; that is,
the total number of occurrences of symbols in $A$. 

Note that if there are many different variables occurring in $A$
then the binary length of $A$ could be as more like $|A|\log |A|$.

{\bf Lemma 1A:} (with Rackoff)
For every nondeterministic TM $M$ there is a polytime
transformation $F_M$ such that for all $x\in\{0,1\}^*$,
$F_M(x)$ is a $\PqR$ formula, and $|F_M(x)| = O(|x|)$, and
$$ A=F_M(x) \mbox{ is satisfiable $\Equiv M$ accepts $x$ in at most
$2^{|x|}$ steps}   $$

{\bf Proof Outline:}
The proof is like that of the Cook-Levin Theorem.
Let $C_0,C_2,...,C_T$ be a computation of
$T=2^n$ steps of $M$ on input $x$, where $n=|x|$.  Here $C_i$
is a bit string of length $O(2^n)$ coding the configuration of
$M$ at step $i$.  Thus the computation can be represented by
a relation $R\subseteq \{0,1\}^{O(n)}$, where $R(\vp,\vq)$
represents bit $\vp$ of $C_{\vq}$.

Then $F_M(x)$ is the prenex form of $S\wedge I\wedge E$ where

$S$ asserts that the computation starts right\\
$I$ asserts that the computation increments right\\
$E$ asserts that  the computation ends right

The formula $E$ is easy, since it merely asserts that the configuration
$C_T$ is in an accepting state.

The formula $S$ asserts that the inital configuration, coded by
$R(\vp, \vec{0})$ (as $\vp$ ranges over all possible values),
represents a tape configuration consisting of
$x$ followed by blanks, and the initial state.

To see how to express this with a formula of length $O(n)$
we assume for simplicity that $x=x_1...x_n$ is a bit string
over $\{0,1\}$.  We show how to construct a formula $S_1(\vp)$
of length $O(n)$
which asserts that for $i=1,...,n$ if $\vp$ represents $i$ in
binary then $(R(\vp,\vec{0})\lra x_i)$.  This explains the
interesting part of the construction of $S$.

To see how to construct $S_1$, let $k= \lceil \log_2 (n+1)\rceil$ and
suppose $p_1,...,p_k$ represent the $k$ low-order bits when
$\vp$ represents a binary number $i$.  (When $i\leq n$, then
the reamaining bits $p_{k+1},...,p_{cn}$ are 0.)
Consider
a Boolean circuit $\alpha$ with inputs $p_1,...,p_k$ and outputs
$r_1,...,r_n$ such that $r_i=1$ iff $p_1...p_k$ represents $i$
in binary.  Note that $\alpha$ can be constructed with $O(n)$ gates
by a simple recursion on $k$. 

Let $\beta(p_1,...,p_k,\vg,\vr)$ be a (quantifier-free) propositional
formula of length $O(n)$
which holds iff the circuit $\alpha$ with input values
$p_1,...,p_k$ takes on values $\vg$ for its internal gates and
values $\vr$ for its output gates.
Then $S_1(\vp)$ is the formula
$$  \forall \vg \forall \vr [(\beta(p_1,...,p_k,\vg,\vr)\wedge Z(\vp)) \ra
 [R(\vp,\vec{0}) \lra ((r_1\wedge x_1)\vee ... \vee (r_n\wedge x_n))]]  $$
where
$$Z(\vp) \equiv \neg p_{k+1}\wedge...\wedge \neg p_{cn}   $$

It remains to discuss the formula $I$.  This asserts that for
all $t< 2^n$, $C_{t+1}$ is the successor configuration to $C_t$
(when $C_t$ and $C_{t+1}$ are represented by $R$.)
Given a reasonable representation of the Turing machine configurations,
it is straightforwward to construct such a $\PqR$-formula $I$ of
length $O(n)$.   $\Box$

{\bf Corollary:}  There is no proof system for the valid
formulas of \QPCR (or for the valid $\Sigma^q_1(R)$ formulas)
with the property that every valid formula $A$ has a proof $P$
such that
\begin{equation}\label{proof}
   |P| = 2^{o(|A|)}  
\end{equation}
where $|P|$ is the bit length of $P$.

{\bf Proof of the Corollary:} 
We use the following

{\bf Fact:}  There is a universal nondeterministic TM $M_0$
such that for every nondeterministic TM $M$ and all sufficiently
large strings $x$ which code $M$,
$$  \mbox{$M_0$ accepts $x$ within $2^{|x|}$ steps $\Equiv$
$M$ accepts $x$ within $2^{0.4|x|}$ steps}  $$

Let $c_0$ be a constant such that, referring to Lemma 1A,
$$  |F_{M_0}(x)|\leq c_0|x|, \mbox{ for all sufficiently long $x$}  $$

Now suppose $\Pi$ is a proof system for unsatisfiability
which violates the Corollary,
so every unsatisfiable $\Sigma^q_1(R)$ formula $A$ has a proof $P$
satisfying (\ref{proof}).
Let $M_1$ be a nondeterministic TM which on input $x$ computes
$A=F_{M_0}(x)$, guesses a proof $P$, and accepts iff $P$ is
a $\Pi$ proof of $A$ (showing that $A$ is unsatisfiable). 
Let $\delta=0.2/c_0$ and
let $x_1$ be a sufficiently long string coding $M_1$.  Then

$M_1$ accepts $x_1$ within $2^{0.4|x_1|}$ steps\\
$\Equiv$ there is a $\Pi$ proof $P$ of $A=F_{M_0}(x_1)$ where 
$|A|\leq c_0|x_1|$ and $|P|\leq 2^{\delta|A|} \leq 2^{0.2|x_1|}$\\
$\Equiv A$ is unsatisfiable\\
$\Equiv M_0$ does not accept $x_1$ within $2^{|x_1|}$ steps\\
$\Equiv M_1$ does not accept $x_1$ within $2^{.4|x_1|}$ steps.

This is a contradiction.  $\Box$

{\bf System G(R):}\\
This is the system $G$ of quantified propositional
calculus described in section 4.6 of Krajicek's book [5], extended
so that formulas are allowed to be \QPCR\ formulas, and we allow
the axiom scheme \AX\ above.  In other words, \GR\
is obtained from \PKR\ by extending the definition of formula,
and allowing the four quantifier rules of \LK (Krajicek, page 58).

{\bf Theorem 4:}  \GR\ is sound and complete.

{\bf Proof:}  Soundness is easy, since as before the axioms are valid
and the rules preserve validity.

We prove that every valid sequent has a \GR\ proof by double induction,
first on the maximum quantifier depth of formulas in the sequent,
and second on the cost $c(S)$ of the sequent, as defined in
Definition 1 above.

To see how to  reduce the quantifier depth, consider the case
$$  \Gamma'\ra\Delta',\exists xA(x)  $$
This can be derived by two applications of $\exists$-{\bf right}
and one of contraction from
$$ \Gamma'\ra\Delta',A(0),A(1)  $$
and this sequent is valid if the previous one is valid.  $\Box$

{\bf Remark:}  It seems that the obvious upper bound for the above
proof length is doubly exponential, even in the case of nonrelativized
$G$, and even for the case nonrelativized ${\bf G}_1$. 

Consider the example 
$$   \ra \exists x_1...\exists x_n(A_1\wedge...\wedge A_m)  $$
If we apply the above method to get rid of the existential quantifiers,
we obtain a sequent with $2^n$ formulas, each of which is a conjunction
of $m$ formulas.  Now to unwind all of these conjunctions in the  usual
way seems to generate $2^{m2^n}$ sequents.  

Rackoff points out that this large upper bound is not surprising
for the relativized case.
In fact, if a simply exponential upper bound could be found, it
would follow from Theorem 3 that \NEXP = {\bf coNEXP}.

However there is a simply exponential upper bound for the nonrelativized
case.

{\bf Theorem 5:}  (See Theorem VII.3.9 in [4].)  Every valid sequent $S$
of \QPC\ (with no $R$) has a tree-like \G\ proof with $O(2^{|S|})$
sequents (not counting weakenings and exchanges), where each sequent has
length $O(|S|)$ and all cut formulas are atomic.

{\bf Work to be done:}

$\bullet$ 
Carry out the translations of the relativized theories
$S^i_2(R)$ and $T^i_2(R)$ into \QPCR.  It may be easier to
translate the two-sorted versions $V^i(R)$ and $TV^i(R)$.
(The theories $V^i$ and $TV^i$ are presented in [4], where
propositional translations are given.)

$\bullet$ Once the translations have been written down, it
should be possible to describe families of valid \QPCR\
formulas corresponding to various search problems, and prove
lower bounds on their \GR\ proof lengths
by the same search problem separations used to  separate various
relativized theories of bounded arithmetic.

{\bf Example:}  Let $\WPHPR$ be a relativized propositional
formula (in fact a $\Sigma^q_2(R)$ formula) representing
the weak pigeonhole principle $\PHPa$ as follows.
(Here we assume that $\vp$ and $\vq$ are vectors of $2n$ variables,
while $\vr$ and $\vs$ are vectors of just $n$ variables.)
$$  \WPHPR \equiv \exists \vp\exists \vq\exists\vr
   [(\vp \not= \vq \wedge R(\vp,\vr)\wedge R(\vq,\vr))
    \vee \forall \vs \neg R(\vp,\vs)]   $$

{\bf Conjecture 1:}  $\langle \WPHPR\rangle$ does not have polysize
$\GGstarR$ proofs.

{\bf Proposed Proof Outline:}  

(i)  Theorem 11.3.1, page 220 of Krajicek's book shows that
the witnessing problem for $\WPHPR$ is not in $FP^{NP(R)}$.

(ii) The witnessing problem for $\GGstarR$ proofs of $\Sigma^q_2(R)$
formulas is in $FP^{NP(R)}$.  This is by analogy with the fact that
the witnessing problem for $\GstR$ proofs of $\Sigma^q_1(R)$ formulas
is in $FP(R)$.

(iii)  If $\langle \WPHPR\rangle$ has polysize $\GGstarR$ proofs,
then given $n$ we could use an $NP$ oracle to find a proof of 
$\WPHPR\rangle$, and then use (ii) to solve the witnessing problem
with an $NP(R)$ oracle.  This contradicts (i).    $\Box$

In the same vein, we know (by translations into bounded depth Frege systems)
that 
$$  S_2(R) \not\vdash PHP(R)   $$
(see Pitassi's thesis).
This suggests 

{\bf Conjecture 2:}  $\langle \PHPR\rangle$ does not have
polysize ${\bf G_i(R)}$-proofs, for any $i$.

Apparently we can translate theorems of $S_2(R)$ both into quasipolysize
families
of bounded depth Frege proofs, and into polysize
families of ${\bf G(R)}$ proofs.  This leads to

{\bf Conjecture 3} (Pudlak):  Find an RSUV style isomormphism between
$AC^0$-Frege and {\bf G(R)}. 

In a slightly different vein, we have

{\bf Conjecture 4:} (Morioka:)  The ITER(R) Tautologies do not
have polysize $\GstR$ proofs.  

{\bf Proposed Proof} (Morioka):  Prove a superpolynomial lower bound for
the circuit size for solving ITER(R).

$\bullet$   Think about using the oracle
separations of NC and P in [2] to separate relativized $\G^*_1${\bf (R)}
and $\G_1${\bf (R)}.

$\bullet$ (Far out:)  Try for lower bounds for unrelativized \G.
Of course there's no super proof system for \QPC\ (including \G)
under the assumption \NP\ $\not=$ {\bf PSPACE}.  Can we get a
lower bound for \G\ proofs under the weaker assumption
{\bf P} $\not=$ {\bf PSPACE}?


{\bf References}

1.  Klaus Aehlig and Arnold Beckmann, Propositional Logic for
Circuit Classes.  CSL 2007.

2. Klaus Aehlig, Stephen Cook, and Phuong Nguyen,
Relativizing Small Complexity Classes and their Theories.
CSL 2007.

3.  Shai Ben-David and Anna Gringauze, On the Existence of Optimal
Propositional Proof Systems and Oracle-Relativized Propositional Logic.
Manuscript, pp 1-12.

4.  Stephen Cook and Phuong Nguyen,  {\em Logical Foundations of Proof
Complexity}.
ASL Perspectives in Logic Series, Cambridge University Press, 2010.

5.  Jan Krajicek, {\em Bounded Arithmetic, Propositional Logic, and
Complexity Theory}.  Cambridge, 1995.

\end{document}